 \documentclass[smallabstract,smallcaptghions]{dccpaper}

\usepackage{epsfig}
\usepackage{citesort}
\usepackage{amsmath}
\usepackage{amssymb}
\usepackage{color}
\usepackage{url}
\usepackage{pgfplots}
\usepackage{caption}
\usepackage{subcaption}
\usepackage{multirow}
\usepgfplotslibrary{groupplots,dateplot}
\newlength{\figurewidth}
\newlength{\smallfigurewidth}

\begin{document}

\title
{\large
\textbf{SLIC: A Learned Image Codec Using Structure and Color}
}

\author{
Srivatsa Prativadibhayankaram$^{\S \dag \star}$, Mahadev Prasad Panda$^{\S \dag \star}$ \thanks{$\star$ Equal contribution}, Thomas Richter$^{\S}$, \\ Heiko Sparenberg$^{\S \P}$, Siegfried Fößel$^{\S}$, Andr\'{e} Kaup$^{\dag}$\\[0.5em]
{\small\begin{minipage}{\linewidth}\begin{center}
 \begin{tabular}{c}
 	$^{\S}$Fraunhofer Institute for Integrated Circuits IIS, Erlangen, 	Germany \\
 	$^{\dagger}$Friedrich-Alexander-Universität Erlangen-Nürnberg, Erlangen, Germany\\
  	$^{\P}$RheinMain University of Applied Sciences, Wiesbaden, Germany \\
 \end{tabular}
\vspace{-2pt}
\begin{tabular}{c c c}
	\url{first.last@iis.fraunhofer.de} &  \url{first.last@fau.de}  & \url{first.last@hs-rm.de}\\
\end{tabular}
\end{center}
\end{minipage}}
}

\maketitle
\thispagestyle{empty}

\begin{abstract}
We propose the structure and color based learned image codec (SLIC) in which the task of compression is split into that of luminance and chrominance. The deep learning model is built with a novel multi-scale architecture for Y and UV channels in the encoder, where the features from various stages are combined to obtain the latent representation. An autoregressive context model is employed for backward adaptation and a hyperprior block for forward adaptation. Various experiments are carried out to study and analyze the performance of the proposed model, and to compare it with other image codecs. We also illustrate the advantages of our method through the visualization of channel impulse responses, latent channels and various ablation studies. The model achieves Bj{\o}ntegaard delta bitrate gains of $7.5\%$ and $4.66\%$ in terms of MS-SSIM and CIEDE2000 metrics with respect to other state-of-the-art reference codecs.
\end{abstract}

\vspace{-5pt}

\Section{Introduction}
\vspace{-5pt}
Development of learned image compression methods has accelerated of late. There are some methods that outperform traditional image codecs such as JPEG \cite{125072} or the intra-frame coding mode of traditional video codecs such as HEVC \cite{6316136} and VVC \cite{9503377}. But the complexity and energy consumption of learned image codecs are many orders of magnitude higher than that of traditional codecs \cite{10222820}. A large number of learned image codecs follow the non-linear transform coding approach introduced in \cite{balleend}. The analysis transform converts an image from the data space to a latent space. This latent representation is quantized to perform irrelevancy reduction and then compressed to a compact form by an entropy coder. The synthesis transform decodes and reconstructs the image. The rate-distortion optimization of such a model can be represented as
\vspace{-5pt}
\begin{equation}
		\mathrm{min}_{\boldsymbol{\theta}, \boldsymbol{\phi}}\{L\}, \text{with  } L(\boldsymbol{\theta}, \boldsymbol{\phi}) = {R}({\boldsymbol{\theta}})   + \lambda \cdot D({\boldsymbol{\theta}, \boldsymbol{\phi}}),
	\label{eqn_rdo}
	\vspace{-18pt}
\end{equation} \\
where $L$ represents the loss term,  $R$ is the rate measured in bits per pixel, $D$ is the distortion term and $\lambda$ is the Lagrangian multiplier. The symbols $\boldsymbol{\theta}$ and $\boldsymbol{\phi}$ indicate the learnable parameters of the analysis and the synthesis transforms respectively. 

Several works targeting various aspects of learned image compression have been developed recently. While some focus on architecture, others develop better context modeling and entropy coding methods. The work in \cite{he2022elic} outperforms many state-of-the-art image codecs including VVC all-intra mode. A codec that makes use of transformers is developed in \cite{9810760}. A novel implicit neural representation based codec is introduced in \cite{10125318}, but the results are not on par with state-of-the-art codecs. A multi-scale skip connection based encoder can be seen in \cite{Zhou_2019_CVPR_Workshops}. The work in \cite{cheng_learned_2020} employs a Gaussian mixture model for better entropy coding, including an autoregressive context model. 

Most learned image codecs operate in the RGB color space. However, there are a few learned codecs that operate in YUV color space \cite{10018070,10222731}. In our prior work \cite{10222731}, we developed a model that contains two branches - one for capturing structure from the luminance or Y channel, and color from chrominance or UV channels. In this work, we adapt the split luma and chroma branches from the color learning model \cite{10222731} to the \emph{Cheng2020} \cite{cheng_learned_2020} model architecture and make various improvements. Firstly, we have a multi-scale encoder block, where features from various stages in the encoder are combined. Secondly, we replace some of the convolutional layers in the hyper synthesis transform by sub-pixel convolution layers that help with better prediction of the latent distribution. Thirdly, we make use of an autoregressive context model, along with an entropy parameter estimation module for backward adaptation, resulting in significant bitrate savings. Finally, instead of the parameter heavy residual attention blocks used in \emph{Cheng2020} \cite{cheng_learned_2020}, we use shuffle attention \cite{shuffleAttention} blocks.

Our main contributions in this work can be outlined as reduction in model complexity with a novel architecture and a better structural as well as color fidelity in reconstruction of images resulting in competitive performance. We illustrate the benefits of our model through various experimental results and ablation studies. We also compare the performance of proposed SLIC model with other codecs -- both traditional and learned, and report our findings.

\begin{figure}[!t]
	\centering
	\includegraphics[width=0.86\textwidth]{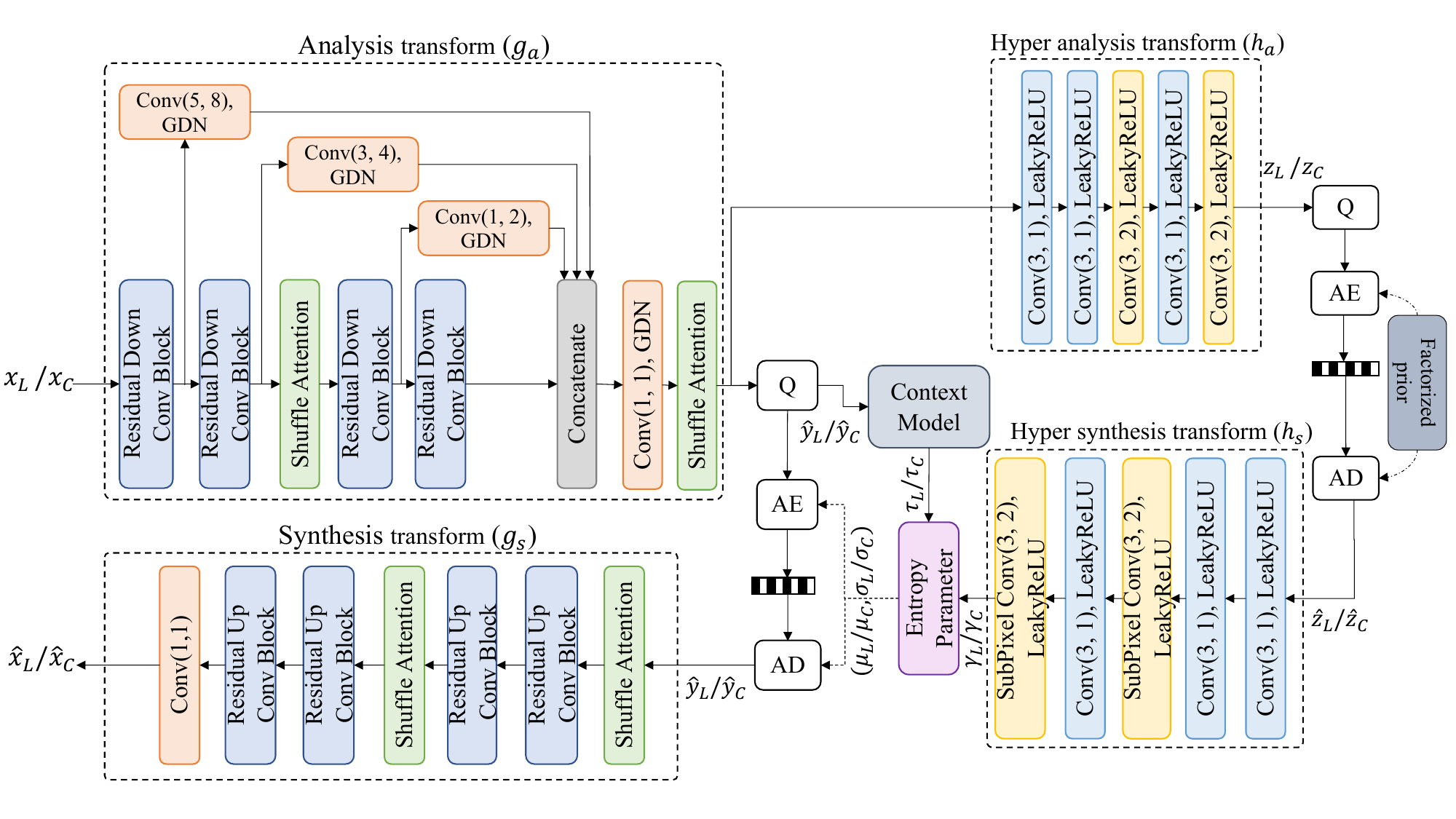}
	\vspace{-5pt}
	\caption{Network architecture of the proposed {SLIC} model. Q represents the quantizer, AE and AD indicate arithmetic encoder and decoder respectively.}
	\label{fig-block}
	\vspace{-15pt}
\end{figure}

\vspace{-2pt}
\Section{Structure and Color Based LIC}\label{sec_method}
\vspace{-8pt}
In this section, we look into the details of the proposed structure and color based learned image codec (SLIC). As mentioned, our model is built based on \cite{cheng_learned_2020} and our prior work in \cite{10222731}. The model has an asymmetric architecture, where the encoder has a higher number of parameters in comparison to the decoder. Additionally, there is an autoregressive context model added to both luminance and chrominance branches. The block diagram is shown in Fig.\ref{fig-block}. It should be noted that all the components are instanced twice - once for luminance (Y) and once for chrominance (UV) channels. \vspace{-12pt}\\ \\
\textbf{Network Architecture:} In the analysis transform blocks, we make use of a multi-scale architecture. The features from various stages of the analysis transform layers are tapped and finally combined. The residual up and down convolution blocks are the same as in \emph{Cheng2020} model. In contrast to \cite{cheng_learned_2020}, we make use of shuffle attention \cite{shuffleAttention} layers instead of the residual attention in both analysis and synthesis transforms. Residual attention consists of $337,536$ parameters in comparison to shuffle attention layer that has only $48$ parameters. It has been experimentally shown in \cite{shuffleAttention} that, shuffle attention layer behaves as a lightweight plug-and-play block, that improves the model performance in various convolutional neural network architectures. In the hyper synthesis transform, we make use of sub-pixel convolution in addition to convolution layers. The sub-pixel convolution is an implementation of deconvolution layer where, a shuffling operation is performed after a standard convolution in low-resolution space. Our autoregressive context block consists of a masked convolution layer with a kernel of size $5\times5$, similar to the model in \cite{cheng_learned_2020}. However, we do not use a Gaussian mixture model for estimating the latent probability distribution. The entropy parameter estimation block consists of three convolutional layers and generates the predicted mean ($\mu$) and scale ($\sigma$) of the latent $\hat{y}$. \vspace{-12pt}\\ \\

\textbf{Loss Function:} As distortion metrics, we use mean squared error (MSE) and multi-scale structural similarity index measure (MS-SSIM) \cite{wang_multiscale_2003} for structural fidelity. Similar to our prior work, we use the color difference metric CIEDE2000 ($\Delta E_{00}^{12}$) \cite{sharma2005ciede2000} to optimize our model for color fidelity. This metric operates in LAB color space with three components, namely luminosity, color, and hue to compute the color difference between two given pixel triplet values. The final loss function based on (\ref{eqn_rdo}), to train the model is :
\vspace{-3pt}
\begin{equation}
\mathrm{min}_{\boldsymbol{\theta}, \boldsymbol{\phi}}\{L\}, \text{with  } L(\boldsymbol{\theta}, \boldsymbol{\phi}) = {R}   + \lambda_{1} \cdot \mathrm{MSE}(\cdot) + \lambda_{2} \cdot (1.0 - \mathrm{MS\text{-}SSIM}(\cdot)) +   \lambda_{3} \cdot \Delta E_{00}^{12}(\cdot),
\label{eqn:loss}
\vspace{-5pt}
\end{equation} \vspace{-12pt}\\
where $\lambda_{1}, \lambda_{2}$, and $\lambda_{3}$ are the Lagrangian multipliers for the metrics MSE, MS-SSIM and CIEDE2000 respectively. It should be noted that MSE and MS-SSIM are estimated in the RGB color space. ${R}$ indicates the total bitrate and consists of four components, namely luma and chroma hyperprior bits, as well as luma and chroma latent bits.

\vspace{-2pt}
\SubSection{Implementation Details}\label{sub_sec_impl}
\vspace{-5pt}
The SLIC model was implemented in Python programming language using \texttt{PyTorch}\footnote[1]{\url{https://pytorch.org}} and \texttt{CompressAI}\footnote[2]{\url{https://github.com/InterDigitalInc/CompressAI}} frameworks. The training data comprised around 118,000 images from the \textsl{COCO2017}\footnote[3]{\url{https://cocodataset.org}} 
training dataset. As validation data, 5,000 randomly chosen images from the ImageNet 
dataset were used. The model was trained for various bitrate configurations with the \textit{ReduceOnPlateau} learning rate scheduler, staring with $1e-4$.  For every configuration, the model was trained for 120 epochs, with images cropped to $256\times256$ and a batch size of 32. The Lagrangian multiplier values were chosen experimentally, based on the range of metric values as: $\lambda_{1}=\{0.001, 0.005, 0.01, 0.02\}$ for MSE, $\lambda_{2}=\{0.01, 0.12, 2.4, 4.8\}$ for MS-SSIM and $\lambda_{3}=\{0.024, 0.12, 0.24, 0.48\}$ for CIEDE2000, similar to \cite{10222731}. Additionally, since the color difference metric CIEDE2000 considers two pixel values, it was modified to work with large batches of image data in the form of tensors efficiently.

The total number of parameters in our SLIC model is around $15$ million, whereas \emph{Cheng2020} model consists of approximately $30$ million parameters in the highest bitrate configuration. In terms of kilo multiply-accumulate operations (kMACs) for each pixel, SLIC needs 829.72 kMAC/pixel and \emph{Cheng2020} model needs 1033.75 kMAC/pixel for an end-to-end forward pass.
\vspace{-2pt}
\SubSection{Model workflow}\label{sub_sec_workflow}
\vspace{-5pt}

The encoder comprises the analysis, hyper analysis and hyper synthesis transform blocks, context model, and the entropy parameter estimation module. The input image $x$ is converted from RGB to YUV color space. The YUV image is split into two components $x_L$ and $x_C$, which are the luminance and chrominance components. The non-linear analysis transform $g_a$ transforms the inputs into the latent representations $y_L$ and $y_C$ respectively. In order to estimate the distribution of the latents, the hyperanalysis transform $h_a$ transforms them into hyperlatents $z_L$ and $z_C$. These hyperlatents are quantized and entropy coded with the factorized prior that is learnt during training. The autoregressive context models generate $\tau_L$ and $\tau_C$ to help encode each latent value and are then combined with the output of hyper synthesis transforms $\gamma_L$ and $\gamma_C$ by the entropy parameter estimation modules to obtain mean and scale values. They are then used to perform arithmetic coding of the quantized latents ${\hat{y}_L}$ and ${\hat{y}_C}$.

The decoder consists of hyperprior blocks, context models, entropy parameter estimation blocks and the synthesis transforms. The decoding process starts with the recovery of hyperlatents $\hat{z}_L$ and $\hat{z}_C$. They are decoded by the hyper synthesis transform $h_s$ to obtain $\gamma_L$ and $\gamma_C$. The context models start with all zeros and iteratively, contexts are estimated for each latent pixel based on the previously decoded latent values. The estimated mean and scale values are then used for entropy decoding and obtain the quantized latents $\hat{y}_L$ and $\hat{y}_C$. Followed by this, they are transformed back into the image space by the synthesis transform $g_s$. The reconstructed luma channel is $\hat{x}_L$ and the chroma component is $\hat{x}_C$. Finally, they are concantenated and converted from YUV to RGB color space, which gives us the final reconstructed image $\hat{x}$.

\vspace{-2pt}
\Section{Experiments and Results}\label{sec_exp}
\vspace{-5pt}
In this section, we provide details about the various experiments that were conducted and illustrate the properties of the proposed SLIC model. We start with the rate-distortion performance and compare SLIC's performance with other methods. Followed by this, we make a visual comparison of image patches reconstructed by various codecs. Then we present visualization of the predicted latent distributions. Finally, we discuss the channel impulse response computed for each latent channel and compare it to that of \emph{Cheng2020} model.

\begin{figure}[!t]
	\centering 
	\resizebox{0.97\textwidth}{!}{% This file was created with tikzplotlib v0.10.1.
\begin{tikzpicture}
	\tikzstyle{every node}=[font=\large]
	\definecolor{darkgray176}{RGB}{176,176,176}
	\definecolor{darkviolet1910191}{RGB}{191,0,191}
	\definecolor{goldenrod1911910}{RGB}{191,191,0}
	\definecolor{green01270}{RGB}{0,127,0}
	\definecolor{lightgray204}{RGB}{204,204,204}
	\definecolor{steelblue31119180}{RGB}{31,119,180}
	
%	\large
	\begin{groupplot}[group style={group size=3 by 1, horizontal sep=2cm}]
		\nextgroupplot[
		tick align=outside,
		tick pos=left,
		x grid style={darkgray176},
		xlabel={Rate (bpp)},
		xmajorgrids,
		xmin=0.00227474618272569, xmax=1.0,
		xtick style={color=black},
		y grid style={darkgray176},
		ylabel={PSNR (dB)   \(\displaystyle \uparrow \)},
		ymajorgrids,
		ymin=25.53222713575, ymax=38,
		ytick style={color=black}
		]
		\addplot [green01270, dash pattern=on 3.7pt off 1.6pt, mark=*, mark size=3, mark options={solid}]
		table {%
			0.098115338 27.98974522
			0.173912113 29.82447267
			0.299725056 31.8515377
			0.456783226 33.87737664
			0.679351191 35.62697935
			0.829 36.654
		};
		\addplot [goldenrod1911910, dash pattern=on 3.7pt off 1.6pt, mark=triangle*, mark size=3, mark options={solid}]
		table {%
			0.132636176215278 27.5610487083333
			0.210235595703125 29.1866699166667
			0.321200900607639 30.931298
			0.479699028862847 32.7869397083333
			0.67002699110243 34.4550285
			0.940165201822917 36.6320425833333
		};
		\addplot [red, dash pattern=on 3.7pt off 1.6pt, mark=triangle*, mark size=3, mark options={solid,rotate=180}]
		table {%
			0.123294406467014 26.8909755833333
			0.189198811848958 28.1997195
			0.288479275173611 29.5900013333333
			0.441050211588542 31.2395902083333
			0.6488037109375 32.9056854583333
			0.967647976345486 35.3012495416667
		};
		\addplot [darkviolet1910191, dash pattern=on 3.7pt off 1.6pt, mark=*, mark size=3, mark options={solid}]
		table {%
			0.0482449 26.14484539
			0.112495422 28.49302175
			0.245816549 31.19987372
			0.490545485 34.26153257
			0.874810113 37.41979316
		};
		\addplot [steelblue31119180, mark=x, mark size=3, mark options={solid}]
		table {%
			0.101962619357639 26.0983017083333
			0.304053412543403 30.0847759166667
			0.607754177517361 32.9184170416667
			0.853075663248698 34.4610417083333
		};
		\addplot [blue, mark=diamond*, mark size=3, mark options={solid}]
		table {%
			0.747585720486111 34.78781825
			0.500698513454861 33.2932780416667
			0.303198920355903 31.3122765416667
			0.0923055013020833 27.4999881666667
		};
		
		\nextgroupplot[
		tick align=outside,
		tick pos=left,
		x grid style={darkgray176},
		xlabel={Rate (bpp)},
		xmajorgrids,
		xmin=0.00227474618272569, xmax=1.0,
		xtick style={color=black},
		y grid style={darkgray176},
		ylabel={MS-SSIM (dB)  \(\displaystyle \uparrow \)},
		ymajorgrids,
		ymin=8.0, ymax=20,
		ytick style={color=black}
		]
		\addplot [green01270, dash pattern=on 3.7pt off 1.6pt, mark=*, mark size=3, mark options={solid}]
		table {%
			0.098115338 10.678
			0.173912113 12.36
			0.299725056 14.567
			0.456783226 16.476
			0.679351191 18.201
			0.829 19.166
		};
		\addplot [goldenrod1911910, dash pattern=on 3.7pt off 1.6pt, mark=triangle*, mark size=3, mark options={solid}]
		table {%
			0.132636176215278 10.8233301010038
			0.210235595703125 12.3639873816379
			0.321200900607639 14.0854223288858
			0.479699028862847 15.9817091981016
			0.67002699110243 17.81696758263
			0.940165201822917 19.6487789593679
		};
		\addplot [red, dash pattern=on 3.7pt off 1.6pt, mark=triangle*, mark size=3, mark options={solid,rotate=180}]
		table {%
			0.123294406467014 10.4523575597074
			0.189198811848958 11.934573295591
			0.288479275173611 13.5046738838869
			0.441050211588542 15.2670919407658
			0.6488037109375 17.1102471554952
			0.967647976345486 19.131352100209
		};
		\addplot [darkviolet1910191, dash pattern=on 3.7pt off 1.6pt, mark=*, mark size=3, mark options={solid}]
		table {%
			0.0482449 8.55762649098114
			0.112495422 10.8455232115259
			0.245816549 13.5761193814512
			0.490545485 16.5912563121794
			0.874810113 19.4988314300131
		};
		\addplot [steelblue31119180, mark=x, mark size=3, mark options={solid}]
		table {%
			0.101962619357639 9.27087805986818
			0.304053412543403 12.9344306012061
			0.607754177517361 17.0634912564401
			0.853075663248698 18.8736502167553
		};
		\addplot [blue, mark=diamond*, mark size=3, mark options={solid}]
		table {%
			0.747585720486111 19.0923095513288
			0.500698513454861 17.1210685025709
			0.303198920355903 14.4531300951154
			0.0923055013020833 10.3767280919996
		};
		
		\nextgroupplot[
		legend cell align={left},
		legend style={
			fill opacity=0.8,
			draw opacity=1,
			text opacity=1,
			at={(1.05,0.5)},
			anchor=west,
			draw=lightgray204
		},
		tick align=outside,
		tick pos=left,
		x grid style={darkgray176},
		xlabel={\large{Rate (bpp)}},
		xmajorgrids,
		xmin=0.00227474618272569, xmax=1.0,
		xtick style={color=black},
		y grid style={darkgray176},
		ylabel={CIEDE2000   \(\displaystyle \downarrow \)},
		ymajorgrids,
		ymin=1.52395847791667, ymax=5.0,
		ytick style={color=black}
		]
		\addplot [green01270, dash pattern=on 3.7pt off 1.6pt, mark=*, mark size=3, mark options={solid}]
		table {%
			0.116923014322917 3.68853575416667
			0.175835503472222 3.13382982916667
			0.270375569661458 2.6595885875
			0.428293863932292 2.2254637375
			0.59661865234375 1.94069311666667
			0.807257758246528 1.69625427916667
		};
		\addlegendentry{{\large Cheng2020}}
		\addplot [goldenrod1911910, dash pattern=on 3.7pt off 1.6pt, mark=triangle*, mark size=3, mark options={solid}]
		table {%
			0.132636176215278 4.3203762125
			0.210235595703125 3.522696675
			0.321200900607639 2.809920875
			0.479699028862847 2.38105271666667
			0.67002699110243 2.04018099166667
			0.940165201822917 1.69037142083333
		};
		\addlegendentry{\large{Hyperprior}}
		\addplot [red, dash pattern=on 3.7pt off 1.6pt, mark=triangle*, mark size=3, mark options={solid,rotate=180}]
		table {%
			0.123294406467014 4.44735250833333
			0.189198811848958 3.71198574166667
			0.288479275173611 3.12720742083333
			0.441050211588542 2.64468957083333
			0.6488037109375 2.26670270833333
			0.967647976345486 1.8638005625
		};
		\addlegendentry{\large{Factorized Prior}}
		\addplot [darkviolet1910191, dash pattern=on 3.7pt off 1.6pt, mark=*, mark size=3, mark options={solid}]
		table {%
			0.0482449 4.313641921
			0.112495422 3.281450525
			0.245816549 2.519429068
			0.490545485 2.162770063
			0.874810113 1.919252791
		};
		\addlegendentry{\large{VTM}}
		\addplot [steelblue31119180, mark=x, mark size=3, mark options={solid}]
		table {%
			0.101962619357639 4.89028457916667
			0.304053412543403 3.09523122083333
			0.607754177517361 2.52140972083333
			0.853075663248698 2.06096312916667
		};
		\addlegendentry{\large{CL (prior)}}
		\addplot [blue, mark=diamond*, mark size=3, mark options={solid}]
		table {%
			0.747585720486111 1.68425972083333
			0.500698513454861 1.97616157916667
			0.303198920355903 2.41886852916667
			0.0923055013020833 3.52604403333333
		};
		\addlegendentry{\large{SLIC (ours)}}
	\end{groupplot}
	
\end{tikzpicture}}
	\vspace{-10pt}
	\caption{RD curves compared with various codecs for the \emph{Kodak} dataset.}
	\label{fig:rd_curves}
	\vspace{-7pt}
\end{figure}
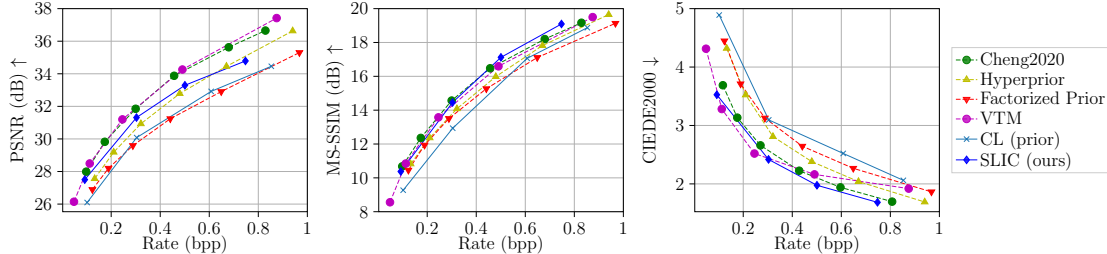

\SubSection{Rate-distortion performance}
\vspace{-5pt}
The model was trained for various bitrate configurations. We measured the rate and distortion values for the Kodak dataset (24 images) spanning various bitrates in the range 0 to 1 bits per pixel (bpp). The distortion metrics PSNR, MS-SSIM, and CIEDE2000 are considered for comparison. The RD values are measured and averaged over all the images for each bitrate configuration. A comparison is made with \emph{Factorized Prior} \cite{balleend}, \emph{Hyperprior} \cite{balle2018variational}, \emph{Cheng2020} \cite{cheng_learned_2020}, \emph{CL model} \cite{10222731}, and VVC reference software VTM\footnote[1]{\url{https://vcgit.hhi.fraunhofer.de/jvet/VVCSoftware_VTM}} \cite{9503377}. The RD curves are shown in Fig. \ref{fig:rd_curves}. Note that for better readability, the MS-SSIM values are converted using $-10 \times log_{10}(1 - \text{MS-SSIM})$ to a decibel (dB) scale.

In terms of PSNR, our model is comparable to the \emph{Hyperprior} model and worse than VTM and \emph{Cheng2020}. But with MS-SSIM curves, our model is comparable to VTM and \emph{Cheng2020} at bitrates less than 0.5 bpp. For the range between 0.5 and 0.8 bpp, we see our model clearly outperforming the rest of the codecs under consideration. Looking at the CIEDE2000 curves, it can be inferred that our model has the best performance at bitrates larger than 0.2 bpp. This highlights the benefit of optimizing the model for color fidelity. 

We also compared the Bj{\o}ntegaard delta bitrate (BD-BR) \cite{bjontegaard2001calculation} and distortion values with VTM as the baseline. The comparison is made for the metrics considered above and reported in Table~\ref{tab:bd_rate}. In terms of PSNR, \emph{Cheng2020} seems to perform the best having $3.4\%$ gain in BD-BR and $0.15$ dB in BD-PSNR values. But in case of MS-SSIM, see a gain of $7.5\%$ in BD-BR and $0.21$ in with BD-MS-SSIM. The BD-BR gain for the proposed SLIC model is the highest for CIEDE2000, with a value of $4.66\%$. This is significantly better when compared to the other codecs.

\begin{table}[!t]
	\centering
	\footnotesize
	\caption{BD-Rate and BD-Distortion comparison with different codecs for the \emph{Kodak} dataset. }
	\label{tab:bd_rate}
	\vspace{-5pt}
	\begin{tabular}{|c|cc|cc|cc|}
		\hline
		\multirow{2}{*}{Codec Name} & \multicolumn{2}{c|}{PSNR}        & \multicolumn{2}{c|}{MS-SSIM}        & \multicolumn{2}{c|}{CIEDE2000}         \\ \cline{2-7} 
		\vspace{1pt}
		& \multicolumn{1}{c|}{\begin{tabular}[c]{@{}c@{}}BD-BR\\ (\%) \end{tabular}} & \begin{tabular}[c]{@{}c@{}}BD-PSNR \\ (dB) \end{tabular} & \multicolumn{1}{c|}{\begin{tabular}[c]{@{}c@{}}BD-BR\\ (\%) \end{tabular}} & \begin{tabular}[c]{@{}c@{}}BD-MS-\\SSIM\end{tabular}  & \multicolumn{1}{c|}{\begin{tabular}[c]{@{}c@{}}BD-BR\\ (\%) \end{tabular}} & \begin{tabular}[c]{@{}c@{}}BD-1\\/CIEDE2000\end{tabular} \\ \hline
		SLIC (Ours)                   & \multicolumn{1}{c|}{\textit{21.74}}   & \textit{-0.83}   & \multicolumn{1}{c|}{\textbf{-7.50}}  & \textbf{0.21}   & \multicolumn{1}{c|}{\textbf{-4.66}}   & \textbf{0.0081} \\
		Cheng2020  \cite{cheng_learned_2020}  & \multicolumn{1}{c|}{\textbf{3.40}}   & \textbf{-0.15}   & \multicolumn{1}{c|}{\textit{-3.32}}  &\textit{ 0.13}   & \multicolumn{1}{c|}{\textit{20.82}}   & \textit{-0.0175} \\
		Hyperprior \cite{balle2018variational}  & \multicolumn{1}{c|}{38.18}   & -1.39   & \multicolumn{1}{c|}{7.91}  & -0.34   & \multicolumn{1}{c|}{67.68}   & -0.0539 \\
		Factorized Prior  \cite{balleend}  & \multicolumn{1}{c|}{78.16}   & -2.35   & \multicolumn{1}{c|}{15.05}  & -0.59   & \multicolumn{1}{c|}{91.62}   & -0.0765 \\
		\hline
	\end{tabular}
	\vspace{-10pt}
\end{table}

\begin{figure}[!t]
	\centering
	\includegraphics[width=0.95\textwidth]{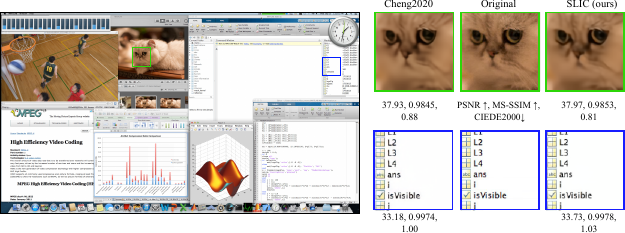}
	\vspace{-12pt}
	\caption{Comparison of reconstructed image patches from SLIC and \emph{Cheng2020}, compressed at a bitrate of around 0.3 bpp. (Best when viewed enlarged on a monitor.)}\label{fig:imgviz}
	\vspace{-15pt}
\end{figure}

\SubSection{Visual comparison of images}
\vspace{-5pt}
The main goal of the split luma and chroma architecture is to optimize for structural and color fidelity. Here we illustrate it with an image for visual comparison of quality. We use the image \texttt{ClassD\_APPLE\_BasketBallScreen\_2560x1440p\_60\_8b\_sRGB.png}, an image composed of natural and synthetic regions, taken from JPEG XL test data. We  compare the decoded images from \emph{Cheng2020} and our SLIC model, compressed at a bitrate of around 0.3 bpp. Two patches of size $128\times128$ in the image are considered, shown in blue and green boxes in Fig. \ref{fig:imgviz}. The quality metrics are provided with the reconstructed patches. Looking at the image crop in blue, which mainly consists of text and icons, it can be seen that the text is reconstructed fairly well by both models. However, on closer inspection, the tiny box with words ‘‘\textit{abc}'' are smudged in the \emph{Cheng2020} image, but are legible in the SLIC image. We also consider a region with natural content indicated by the green box, which is a cat face. Here we observe that the highly textured regions are smoothed in both cases. But the complex textures are better preserved by our model in comparison to \emph{Cheng2020}.

\begin{figure}[!t]
	\centering
	\includegraphics[width=\textwidth]{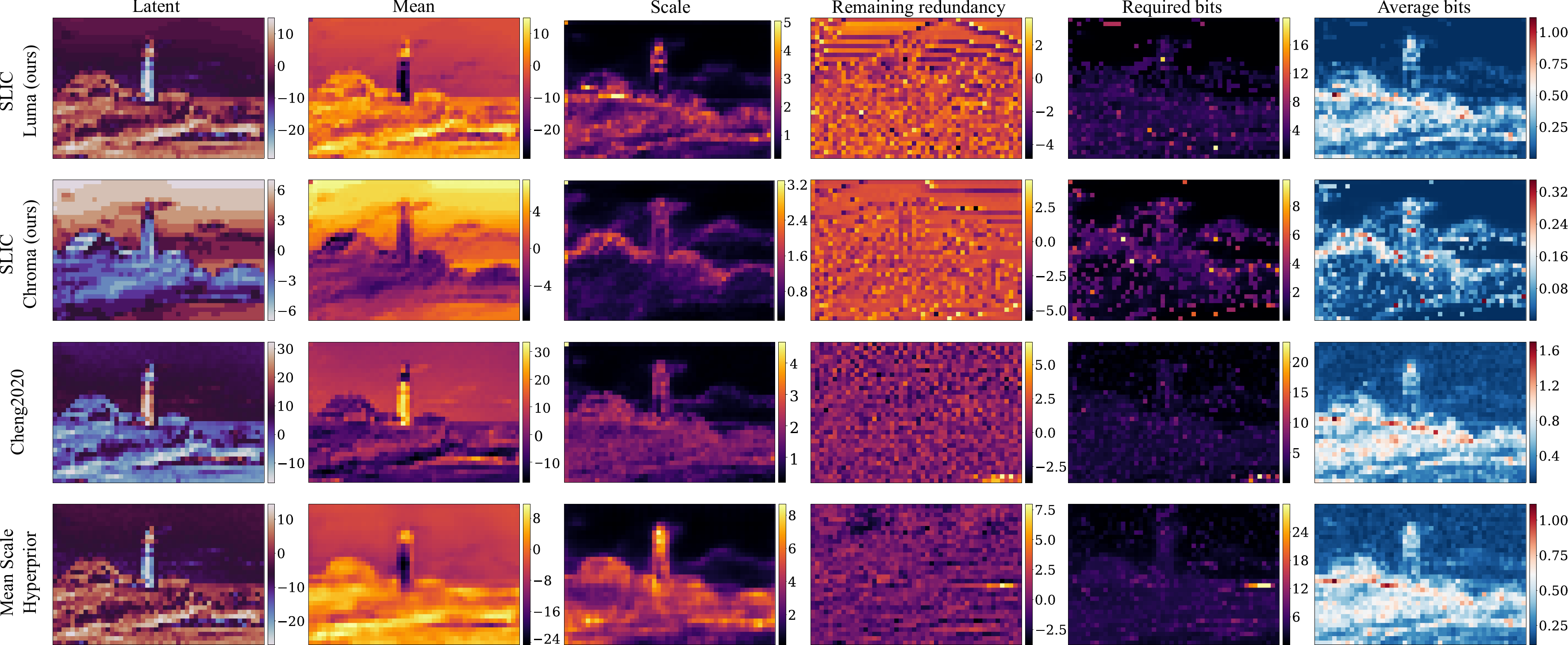}
	\vspace{-15pt}
	\caption{Latent visualization of proposed SLIC, \emph{Cheng2020}\cite{cheng_learned_2020}, and \emph{Hyperprior} \cite{balle2018variational} models for the image \texttt{kodim21.png}. (Best when viewed enlarged on a monitor.)}\label{fig_latent}
	\vspace{-15pt}
\end{figure}

\vspace{-2pt}
\SubSection{Visualization of predicted distributions of latents}
\vspace{-5pt}
Similar to the visualization in \cite{cheng_learned_2020}, we have illustrated the effect of different entropy models in Fig. \ref{fig_latent}. We used the \texttt{kodim21.png} from Kodak dataset as a test image. Here we visualize the latent channels and entropy of the proposed SLIC, \emph{Hyperprior} \cite{balle2018variational}, and that of \emph{Cheng2020} \cite{cheng_learned_2020} models depicted in each row. The most contributing latent channel in terms of bitrate, or in other words the channel with highest entropy is visualized for each codec. The first two rows represent the luma and chroma branches of our SLIC model. The \emph{Cheng2020} results are shown in the third row. The fourth row consists of results from the \emph{Hyperprior} \cite{balle2018variational} model with mean and scale hyperprior.

The latent channel ($\hat{y}$) for each codec is visualized in the first column. The predicted mean $\mu$ and variance $\sigma$ values for the latent channel are shown in the second and third columns respectively. We see that the predicted mean $\mu$ has structure similar to the latent $\hat{y}$. The regions not captured by the predicted mean, appear in the visualization of scale $\sigma$, shown in column 3. The scale visualization shows lower values at smoother regions and higher values at edges and highly textured areas. It can be clearly observed, that our model, as well as \emph{Cheng2020} have sparse and lower values in the scale visualization in column 3. But they are higher and denser for the \emph{Hyperprior} model. This can be attributed to the causal context modeling used in both SLIC and \emph{Cheng2020} models. 

The normalized values representing the remaining redundancy not captured by the mean or scale predictions are visualized in the fourth column. Their values are measured by $\frac{\hat{y} - \mu}{\sigma} $. The required bits for encoding each pixel in the latent channel is computed as $-log_{2}( p_{\hat{y}|\hat{z}}(\hat{y}|\hat{z}))$ using the predicted probability distribution and visualized in the fifth column, where $\hat{z}$ represents the decoded hyperlatent. It provides an insight into the number of bits required to encode the remaining redundancy. Lower redundancy enables lesser number of bits for coding. Finally, the average number of bits required per channel for each latent pixel, shown in the last column is computed using $-\frac{1}{N}\sum_{i} log_{2}( p_{\hat{y}_{i}|\hat{z}_{i}}(\hat{y}_{i}|\hat{z}_{i}))$ where, $i = \{0, 1, ..., N-1\}$ and $N$ is the number of latent channels. 

Although column 5 gives an overview with regards to the required bits, it is specific to the channel with the highest entropy. In order to get a complete picture , we compute the average bits for each latent pixel. In column 6, it can be seen that more bits are required to encode highly textured regions. We observe that structured regions require higher number of bits in the luma part, shown in row 1. However, the regions with large change in color values need more bits in the chroma component, as seen in row 2. The benefits of encoding the luma and chroma latents individually with separate entropy models can thus be seen. We observe this behavior with all bitrate configurations.

\begin{figure}[!t]\label{fig_imp}
	\centering
	\includegraphics[width=0.60\textwidth]{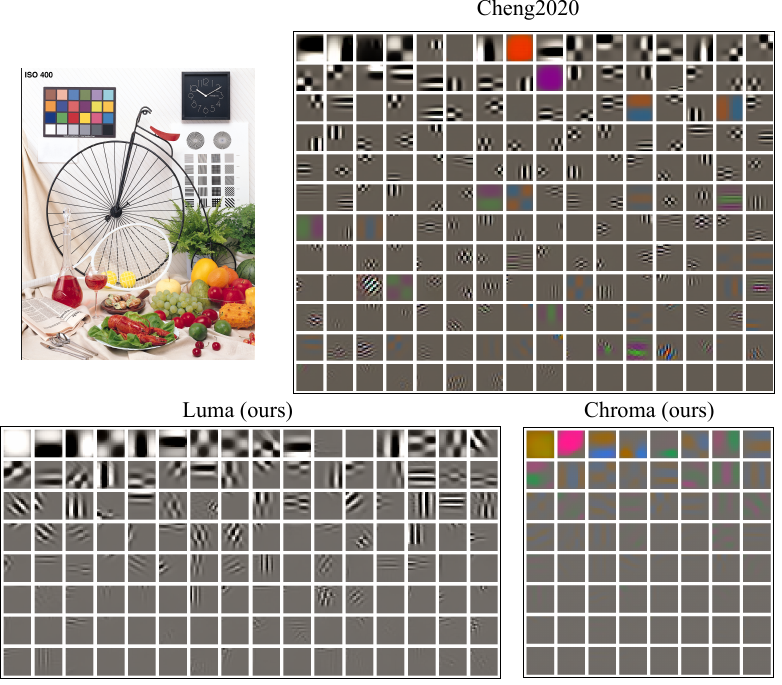}
	\vspace{-5pt}
	\caption{Impulse responses of image \texttt{ClassA\_8bit\_BIKE\_2048x2560\_8b\_RGB.png}.}
	\vspace{-15pt}
\end{figure}

\vspace{-2pt}
\SubSection{Channel Impulse Response}
\vspace{-5pt}
Based on our prior work \cite{10222731}, we compute the channel impulse response of the proposed SLIC model and compare it with that of \emph{Cheng2020}. The channels are sorted in decreasing order of their bitrate contributions, measured using $R_{n} = -log_{2}(p_{n})$ for a channel $n$ using the prior probability $p_n$. The low frequency components appear first, followed by the higher frequencies. We observe a mixture of color and structure in the impulse response of the \emph{Cheng2020} model. Whereas, in our model we have a separation into structure in luminance and color in chrominance components. For luma channel impulse response, we see similarity with linear orthogonal transforms such as discrete cosine transform (DCT).

\vspace{-5pt}
\Section{Ablation Study}\label{sec_ablation}
\vspace{-10pt}
We report two ablation studies on our model. Firstly, we studied the effect of various loss functions on the model performance. Secondly, we evaluated variants of the context model. For all the experiments, we used the same model architecture and training environment as described in the previous section, unless stated. The Kodak dataset was used for evaluating the experiments.

\begin{figure}[t!]
	\centering
	\resizebox{0.83\textwidth}{!}{% This file was created with tikzplotlib v0.10.1.
\begin{tikzpicture}
	
	\definecolor{darkgray176}{RGB}{176,176,176}
	\definecolor{green01270}{RGB}{0,127,0}
	\definecolor{lightgray204}{RGB}{204,204,204}
	\tikzstyle{every node}=[font=\normalsize]
	
	\begin{groupplot}[group style={group size=3 by 1, horizontal sep=2cm}]
		\nextgroupplot[
		legend cell align={left},
		legend style={
			fill opacity=0.8,
			draw opacity=1,
			text opacity=1,
			at={(0.97,0.03)},
			anchor=south east,
			draw=lightgray204
		},
		tick align=outside,
		tick pos=left,
		x grid style={darkgray176},
		xlabel={Rate (bpp)},
		xmajorgrids,
		xmin=0.0419567532009547, xmax=0.781187099880642,
		xtick style={color=black},
		y grid style={darkgray176},
		ylabel={PSNR (dB)   \(\displaystyle \uparrow \)},
		ymajorgrids,
		ymin=26.3664945875, ymax=35.1888336625,
		ytick style={color=black}
		]
		\addplot [green01270, mark=*, mark size=3, mark options={solid}]
		table {%
			0.653927273220486 34.489889625
			0.361948649088542 32.108249625
			0.075558132595486 26.76751
		};
		\addlegendentry{MSE}
		\addplot [blue, mark=triangle*, mark size=3, mark options={solid}]
		table {%
			0.711341010199653 34.6839243333333
			0.470679389105903 33.0507340833333
			0.302341037326389 31.3391520833333
			0.0912780761718749 27.4802987916667
		};
		\addlegendentry{MSE+CIEDE2000}
		\addplot [red, mark=triangle*, mark size=3, mark options={solid,rotate=180}]
		table {%
			0.747585720486111 34.78781825
			0.500698513454861 33.2932780416667
			0.303198920355903 31.3122765416667
			0.0923055013020833 27.4999881666667
		};
		\addlegendentry{MSE+CIEDE2000+MS-SSIM}
		
		\nextgroupplot[
		tick align=outside,
		tick pos=left,
		x grid style={darkgray176},
		xlabel={Rate (bpp)},
		xmajorgrids,
		xmin=0.0419567532009547, xmax=0.781187099880642,
		xtick style={color=black},
		y grid style={darkgray176},
		ylabel={MS-SSIM (dB)   \(\displaystyle \uparrow \)},
		ymajorgrids,
		ymin=9.30639543458333, ymax=19.7068350904167,
		ytick style={color=black}
		]
		\addplot [green01270, mark=*, mark size=3, mark options={solid}]
		table {%
			0.653927273220486 17.9811702916667
			0.361948649088542 15.2323021666667
			0.075558132595486 9.77914269166666
		};
		\addplot [blue, mark=triangle*, mark size=3, mark options={solid}]
		table {%
			0.711341010199653 18.699266125
			0.470679389105903 16.5925230416667
			0.302341037326389 14.5948828333333
			0.0912780761718749 10.5479184583333
		};
		\addplot [red, mark=triangle*, mark size=3, mark options={solid,rotate=180}]
		table {%
			0.747585720486111 19.2340878333333
			0.500698513454861 17.2873269166667
			0.303198920355903 14.6540652916667
			0.0923055013020833 10.6018311875
		};
		
		\nextgroupplot[
		tick align=outside,
		tick pos=left,
		x grid style={darkgray176},
		xlabel={Rate (bpp)},
		xmajorgrids,
		xmin=0.0419567532009547, xmax=0.781187099880642,
		xtick style={color=black},
		y grid style={darkgray176},
		ylabel={CIEDE2000   \(\displaystyle \downarrow \)},
		ymajorgrids,
		ymin=1.536141268125, ymax=4.79474722770833,
		ytick style={color=black}
		]
		\addplot [green01270, mark=*, mark size=3, mark options={solid}]
		table {%
			0.653927273220486 1.9749014375
			0.361948649088542 2.48525432083333
			0.075558132595486 4.646628775
		};
		\addplot [blue, mark=triangle*, mark size=3, mark options={solid}]
		table {%
			0.711341010199653 1.71005406666667
			0.470679389105903 2.02073635416667
			0.302341037326389 2.38906715833333
			0.0912780761718749 3.58778780833333
		};
		\addplot [red, mark=triangle*, mark size=3, mark options={solid,rotate=180}]
		table {%
			0.747585720486111 1.68425972083333
			0.500698513454861 1.97616157916667
			0.303198920355903 2.41886852916667
			0.0923055013020833 3.52604403333333
		};
	\end{groupplot}
	
\end{tikzpicture}}
	\vspace{-5pt}
	\caption{RD performance for different loss functions.}
	\vspace{-5pt}
	\label{ablationLoss}
\end{figure}
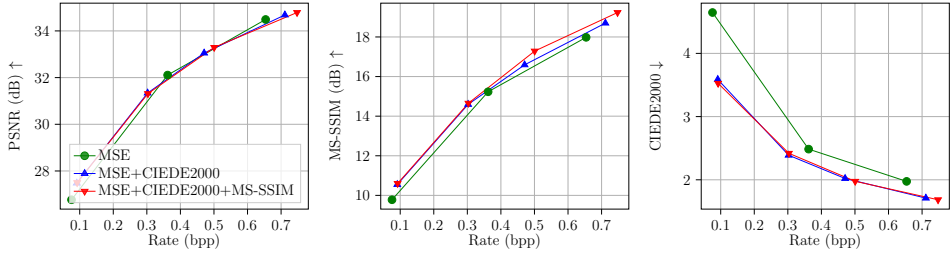

\SubSection{Effect of loss function}
\vspace{-5pt}
We initially trained our model with the MSE distortion metric. Followed by this, we trained our model with a combination of MSE and CIEDE2000 metrics. Finally, we trained the model with MSE, MS-SSIM, and CIEDE2000 metrics, as in (\ref{eqn:loss}). The findings on the effect of loss function on RD performance is shown in Fig. \ref{ablationLoss}. It can be observed that using the color difference metric in the loss function not only improves the color fidelity, but also the structural fidelity. This is evident from the MS-SSIM curves. However, using MS-SSIM in addition to the other two metrics, further improves the performance. Having MS-SSIM additionally in the loss term does not seem to impact PSNR or CIEDE2000 values.

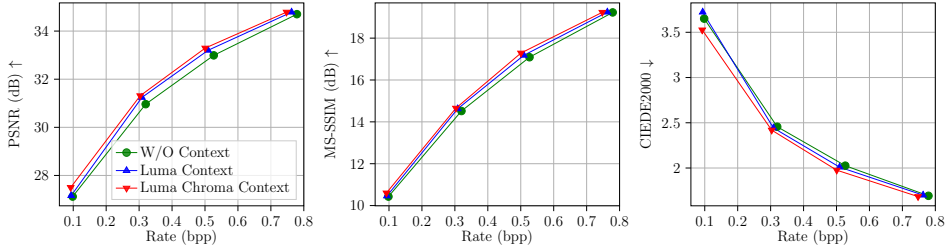
\begin{figure}[!t]
	\centering
	\resizebox{0.83\textwidth}{!}{% This file was created with tikzplotlib v0.10.1.
\begin{tikzpicture}

\definecolor{darkgray176}{RGB}{176,176,176}
\definecolor{green01270}{RGB}{0,127,0}
\definecolor{lightgray204}{RGB}{204,204,204}

\begin{groupplot}[group style={group size=3 by 1, horizontal sep=2cm}]
\nextgroupplot[
legend cell align={left},
legend style={
  fill opacity=0.8,
  draw opacity=1,
  text opacity=1,
  at={(0.97,0.03)},
  anchor=south east,
  draw=lightgray204
},
tick align=outside,
tick pos=left,
x grid style={darkgray176},
xlabel={Rate (bpp)},
xmajorgrids,
xmin=0.0579659356011284, xmax=0.8,
xtick style={color=black},
y grid style={darkgray176},
ylabel={PSNR (dB)   \(\displaystyle \uparrow \)},
ymajorgrids,
ymin=26.73848259375, ymax=35.1711199479167,
ytick style={color=black}
]
\addplot [green01270, mark=*, mark size=3, mark options={solid}]
table {%
0.77909681532118 34.7025051666667
0.526540120442708 32.9907753333333
0.319959852430556 30.9627911666667
0.0982598198784722 27.1217842916667
};
\addlegendentry{W/O Context}
\addplot [blue, mark=triangle*, mark size=3, mark options={solid}]
table {%
0.762875027126736 34.7863278333333
0.509423149956597 33.201772375
0.308573404947917 31.2425003333333
0.093777126736111 27.162267
};
\addlegendentry{Luma Context}
\addplot [red, mark=triangle*, mark size=3, mark options={solid,rotate=180}]
table {%
0.747585720486111 34.78781825
0.500698513454861 33.2932780416667
0.303198920355903 31.3122765416667
0.0923055013020833 27.4999881666667
};
\addlegendentry{Luma Chroma Context}

\nextgroupplot[
tick align=outside,
tick pos=left,
x grid style={darkgray176},
xlabel={Rate (bpp)},
xmajorgrids,
xmin=0.0579659356011284, xmax=0.8,
xtick style={color=black},
y grid style={darkgray176},
ylabel={MS-SSIM (dB)   \(\displaystyle \uparrow \)},
ymajorgrids,
ymin=9.98881964375, ymax=19.6833483979167,
ytick style={color=black}
]
\addplot [green01270, mark=*, mark size=3, mark options={solid}]
table {%
0.77909681532118 19.2282449583333
0.526540120442708 17.087869125
0.319959852430556 14.5173724166667
0.0982598198784722 10.4294800416667
};
\addplot [blue, mark=triangle*, mark size=3, mark options={solid}]
table {%
0.762875027126736 19.242688
0.509423149956597 17.180273625
0.308573404947917 14.6239965625
0.093777126736111 10.4605839375
};
\addplot [red, mark=triangle*, mark size=3, mark options={solid,rotate=180}]
table {%
0.747585720486111 19.2340878333333
0.500698513454861 17.2873269166667
0.303198920355903 14.6540652916667
0.0923055013020833 10.6018311875
};

\nextgroupplot[
tick align=outside,
tick pos=left,
x grid style={darkgray176},
xlabel={Rate (bpp)},
xmajorgrids,
xmin=0.0579659356011284, xmax=0.8,
xtick style={color=black},
y grid style={darkgray176},
ylabel={CIEDE2000    \(\displaystyle \downarrow \)},
ymajorgrids,
ymin=1.582331104375, ymax=3.82476066645833,
ytick style={color=black}
]
\addplot [green01270, mark=*, mark size=3, mark options={solid}]
table {%
0.77909681532118 1.692717575
0.526540120442708 2.02584502916667
0.319959852430556 2.45619502916667
0.0982598198784722 3.65082870833333
};
\addplot [blue, mark=triangle*, mark size=3, mark options={solid}]
table {%
0.762875027126736 1.70027925
0.509423149956597 2.009602375
0.308573404947917 2.44267642916667
0.093777126736111 3.72283205
};
\addplot [red, mark=triangle*, mark size=3, mark options={solid,rotate=180}]
table {%
0.747585720486111 1.68425972083333
0.500698513454861 1.97616157916667
0.303198920355903 2.41886852916667
0.0923055013020833 3.52604403333333
};
\end{groupplot}

\end{tikzpicture}}
	\vspace{-5pt}
	\caption{RD performance for different configurations of context model.}
	\label{ablationContext}
	\vspace{-5pt}
\end{figure}

\SubSection{Effect of Context Model}
\vspace{-5pt}
We compare three configurations of the SLIC model, namely without context model, context block in luma branch only, and context in both luma and chroma branches. A total of 12 models (four per variant) were trained. We report the RD performance in Fig. \ref{ablationContext}. It can be observed that adding context improves performance in all three metrics, due to the backward adaptation, where predictions are based on a causal context. The third variant with context model in both branches performs the best. However, with the context modeling blocks, additional time is required to encode and decode, due to the causal nature of context modeling. Table.\ref{bd_rate2} lists the comparison of BD-Rate and BD-Distortion values made with VTM as the baseline. It shows that the context model in both luma and chroma branches provides the most gains.

\begin{table}[!t]
	\centering
	\footnotesize
	\caption{BD-Rate and BD-Distortion comparison with different codecs. }
	\vspace{-3pt}
	\label{bd_rate2}
	\begin{tabular}{|c|cc|cc|cc|}
		\hline
		\multirow{2}{*}{Configuration} & \multicolumn{2}{c|}{PSNR}        & \multicolumn{2}{c|}{MS-SSIM}        & \multicolumn{2}{c|}{CIEDE2000}         \\ \cline{2-7} 
		& \multicolumn{1}{c|}{\begin{tabular}[c]{@{}c@{}}BD-BR\\ (\%) \end{tabular}} & \begin{tabular}[c]{@{}c@{}}BD-\\PSNR (dB) \end{tabular} & \multicolumn{1}{c|}{\begin{tabular}[c]{@{}c@{}}BD-BR\\ (\%) \end{tabular}} & \begin{tabular}[c]{@{}c@{}}BD-MS-\\SSIM\end{tabular}  & \multicolumn{1}{c|}{\begin{tabular}[c]{@{}c@{}}BD-BR\\ (\%) \end{tabular}} & \begin{tabular}[c]{@{}c@{}}BD-1\\/CIEDE2000\end{tabular} \\ \hline
		W/O Context                   & \multicolumn{1}{c|}{39.89}   & -1.345   & \multicolumn{1}{c|}{0.68}  &  -0.098   & \multicolumn{1}{c|}{6.96}   & -0.002 \\
		Luma Context  & \multicolumn{1}{c|}{27.90}   & -0.997    & \multicolumn{1}{c|}{-4.53}  & 0.115   & \multicolumn{1}{c|}{3.32}   &  0.001 \\
		Luma Chroma Context  & \multicolumn{1}{c|}{\textbf{21.74}}   & \textbf{-0.827}   & \multicolumn{1}{c|}{\textbf{-7.50}}  & \textbf{0.205}   & \multicolumn{1}{c|}{\textbf{-4.66}}   & \textbf{0.008} \\
		\hline
	\end{tabular}
	\vspace{-13pt}
\end{table}

\vspace{-5pt}
\Section{Conclusion}\label{sec_conclusion}
\vspace{-5pt}
A learned image codec that uses structure and color separately, called SLIC is proposed. We show that splitting the image compression task based on luminance and chrominance components not only improves performance, but also reduces the model complexity significantly. The asymmetric architecture makes for more practical image compression, with BD-BR gains of 7.5\% for MS-SSIM. Although we outperform various codecs in terms of MS-SSIM and CIEDE2000, we still lack in terms of PSNR, which we plan to address in a future work. As continuation of this work, we plan to speed up context modeling through parallelization and also compare with other learned image codecs operating in YUV color space, such as JPEG AI. 

\vspace{-5pt}
\Section{References}\label{sec_ref}
\vspace{-5pt}
{\tiny \bibliographystyle{IEEEbib}}
\bibliography{refs}

\end{document}